\documentclass[aps,prl,showpacs,twocolumn,superscriptaddress]{revtex4-1}
\usepackage{amsmath}
\usepackage{amsfonts}
\usepackage{braket}
\usepackage{amssymb}
\usepackage{graphicx}
\usepackage{natbib}
\usepackage[colorlinks=true,linkcolor=blue,urlcolor=black,citecolor=blue]{hyperref}

\newcommand{\diag}{\mathop{\mathrm{diag}}}

\begin{document}
\title{Ground state cooling of quantum systems via a one-shot measurement}

\author{P. V. Pyshkin}
\affiliation{Beijing Computational Science Research Center, Beijing 100084, China}
\affiliation{Department of Theoretical Physics and History of Science, The Basque Country University (EHU/UPV), PO Box 644, 48080 Bilbao, Spain}
\affiliation{Ikerbasque, Basque Foundation for Science, 48011 Bilbao, Spain}

\author{Da-Wei Luo}
\affiliation{Beijing Computational Science Research Center, Beijing 100084, China}
\affiliation{Department of Theoretical Physics and History of Science, The Basque Country University (EHU/UPV), PO Box 644, 48080 Bilbao, Spain}
\affiliation{Ikerbasque, Basque Foundation for Science, 48011 Bilbao, Spain}

\author{J. Q. You}
\affiliation{Beijing Computational Science Research Center, Beijing 100084, China}

\author{Lian-Ao Wu}
\email{lianaowu@gmail.com}
\affiliation{Department of Theoretical Physics and History of Science, The Basque Country University (EHU/UPV), PO Box 644, 48080 Bilbao, Spain}
\affiliation{Ikerbasque, Basque Foundation for Science, 48011 Bilbao, Spain}

\date{\today}

\begin{abstract}
We prove that there exists a family of quantum systems that can be cooled to their ground states by a one-shot projective measurement on the ancillas coupled to these systems.  Consequently, this proof gives rise to the conditions for achieving the one-shot measurement ground-state cooling (OSMGSC).  We also propose a general procedure for finding unitary propagators and corresponding Hamiltonians to realize such cooling by means of inverse engineering technique. 
\end{abstract}
%Quantum mechanics, 03.65.-w
%Quantum state engineering and measurements, 42.50.Dv 
%Cooling of atoms, ions, and molecules, 37.10.De

\maketitle

{\it Introduction.}{\bf--} Quantum ground-state cooling of  small objects exemplified by nano-systems has long been a challenge and one of the most desirable quantum technologies. Physically, the cooling process can be formulated as a transformation from an initial thermal state of a small object into its ground state. The transformation is irreversible and cannot be realized when the object is isolated. It is an indispensable part in the initialization of quantum devices such as an adiabatic quantum computer~\cite{Childs2001,Farhi2001,Sarandy2005,Hammerer2009,You2011}, and it also plays a crucial role in the ultrahigh-precision measurements using mechanical resonators~\cite{Bocko1996,Caves1980,Li2011}. Over the years,  scientists have been under great efforts to ground-state cooling techniques~\cite{Li2011,sbc_Diedrich1989,sbc_Metzger2004,sbc_Neuhauser1978,Xu2014,Zhang2005,Blake2012,Ketterle1992,Wilson-Rae2004,Wilson-Rae2007,Wu2013,Zippilli2005,sbc_Rocheleau2010}, in particular on the sideband cooling~\cite{sbc_Diedrich1989,sbc_Metzger2004,sbc_Neuhauser1978}.% the bang-bang cooling through a Cooper pair box~\cite{Zhang2005} and single-shot state-swapping cooling via a superconductor~\cite{Wang2011}.

Recently, the ground-state cooling of small objects via quantum measurements has been proposed theoretically~\cite{Li2011} and verified experimentally~\cite{Xu2014}. In this approach, the target system $A$ is coupled to an ancilla $B$. The composite system $A+B$ undergoes a unitary evolution for a {\em random} interval of time before a projective measurement is taken on the ancilla. %discarding all measurements whose outcome is not the ground state of the ancilla.
The evolution-measurement procedure is then repeated, if the outcome of this projective measurement is found to be the ground state. It was reported that efficient ground-state cooling can be achieved by repeating such random-time-interval evolutions and measurements, and the cooling efficiency hardly depends on time intervals between any two consecutive measurements~\cite{Li2011} but increases with the frequency of measurements. %becoming exact when $N$ being infinity. 
The major disadvantage of this cooling approach is that it requires many measurements to achieve ground cooling, and consequently the survival probability becomes so small that a very large ensemble of identical systems is required. Here we prove an existence theorem that, for a family of physical systems, guarantees ground-state cooling by making a {\em{one-shot projective measurement}} at a specified time, %It relaxes excessive requirements such as a large ensemble of identical systems.
and derive explicit conditions for this one-shot measurement cooling method to be valid. Furthermore, we show a general approach to engineering Hamiltonians that are able to realize the one-shot measurement cooling %by designing the evolution operator satisfying the one-measurement cooling condition, and the corresponding physical Hamiltonian can be found 
by means of inverse engineering techniques. For existing Hamiltonians, our approach can be used to find the optimal times when the projective measurement is taken, and it is interesting to note that the probability to realize the one-shot measurement cooling remains high even if the above-mentioned conditions are not strictly satisfied.

{\it Existence proof of one-shot measurement ground-state cooling} (OSMGSC).{\bf--}
Consider an $n$-level quantum system $A$ coupled with an ancilla $B$ with $m$ levels.  The system $A$ may not be experimentally accessible such that ground-state cooling cannot be processed. To remedy this,  we choose a controllable ancilla $B$ to manipulate the system $A$ through the $A$-$B$ interaction and measurements on $B$. Without loss of generality, we prepare the ancilla in its ground state $\ket g$. The joint unitary propagator for the composite system $A+B$ is expressed by $U(t)=\sum_{i,\alpha;j,\beta}U_{i,\alpha;j,\beta}(t)|i\rangle_A\langle j|\otimes|\alpha\rangle_B\langle \beta|$ with $nm\times nm$ independent real parameters. It has been shown that it is possible to achieve efficient ground-state cooling with repeated projective measurements. We now devise a new protocol that allows ground-state cooling by making a one-shot projective measurement on the ancilla $B$ after the composite system $A+B$ undergoes a joint unitary evolution for a specified time interval.

\begin{figure}
\begin{center}
\includegraphics[scale=.9]{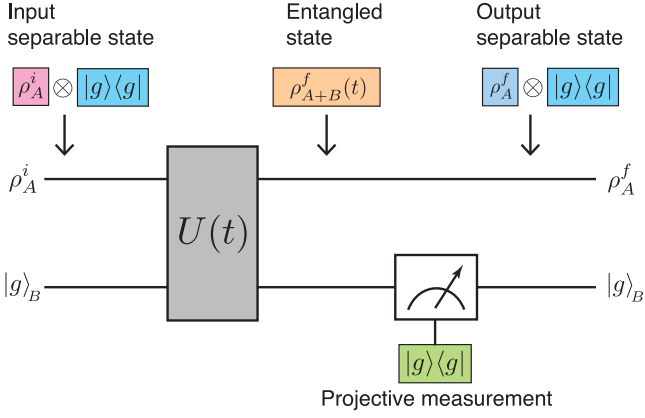}
\end{center}
\caption{One-shot measurement ground-state cooling scheme. An ancilla $B$ initially in its ground state is coupled to the target $A$ that is expected to be cooled to the ground state. The composite system undergoes a unitary evolution  for a duration of $t$ and the final state $\rho_{A+B}^f(t)$ is entangled. A projective measurement is then applied on the ancilla.  We discard the result if the output of the ancilla is not the ground state.}
\label{pic0}
\end{figure}

%We now prove that there exists a family of quantum systems that can be cooled efficiently to their ground states by a single projective measurement on the coupled ancillary system. 

Consider a general initial state of system $A$,
\begin{equation} \label{psiinit}
\rho^i_A=\sum_{l=0}^{n}p_l\ket{l}\bra{l},
\end{equation}
where $p_l$ are probabilities arising from the thermal bath, and $\ket{l}$ are energy eigenstates of $A$ and $n>1$ is the dimension of the system Hilbert space. After the joint unitary evolution, the state becomes $\rho_{A+B}^f(t)=U \rho^i_A\otimes |g \rangle\langle g| U ^\dagger$. We then make a projective measurement on $B$.  Given that the outcome of the measurement is $|g \rangle$,  the output of the composite $A+B$ is $\rho_A^f(t)\otimes|g\rangle\langle g|$, where $\rho_A^f(t)$ is the final state of the target $A$. A schematic diagram of the cooling procedure is depicted in Fig.~\ref{pic0}. 

After the projective measurement, the probability $P_{l,g}$ of finding the composite system $\rho_{A+B}^f(t)$ in state $\ket{l,g}$ is
\begin{equation} \label{p_n_g}
P_{l,g} =\sum_{k}p_k |U_{l,g;k,g}|^2,
\end{equation}
It manifests that the conditions to achieve ground-state cooling are all $P_{l,g}=0$ for $l\geq 1$ such that there is no population in the excited states. Seeing that all $p_k\ge 0$, if we require stronger constraints
\begin{equation} \label{mainreq}
U_{l,g;k,g}\equiv 0, \; l\geq1, \forall k,
\end{equation}
the conditions $P_{l,g}=0$  will always hold true. 
The equation provides $2n^2-2n$ constraints upon the $nm\times nm$ real parameters of the propagator $U$, leaving $n\left( n(m^2-2)+2 \right)$ free parameters. This clearly shows that there always exists a propagator $U$ such that the system can be cooled to its ground states by a one-shot measurement.

We now prove that one can always construct a Hamiltonian $H$ which drives the propagator $U(t)$ to satisfy Eq.~\eqref{mainreq}.  Generally, a Hamiltonian can be expressed in terms of the propagator 
\begin{equation} \label{hamilt}
H = i\dot U U^\dagger.
\end{equation}
By imposing the hermiticity of the Hamiltonian, it can be shown that there are $p=n(m-1)\left(2n(m+1)-1\right)$ free parameters to choose in the construction of Hamiltonian $H$. Since the number of free parameter $p>1$ for any $n\ge 1$ and $m\ge 2$, it is self-evident that there always exists a family of Hamiltonians $H$ which can be used to realize the ground-state cooling by a one-shot measurement.

The Hamiltonians of these {\em good for cooling} systems can be found by using Eqs.~\eqref{mainreq} and~\eqref{hamilt}. Note that $p$ is even greater than~$nm\times nm$, the number of real parameters of a hermitian matrix $H$. As such, one can always construct a Hamiltonian from the unitary propagator by using inverse engineering control technique~\cite{inv_Jing2013}.%rather than directly solving for an optimal measurement time for some given Hamiltonian.

{\it Inverse engineering.}{\bf--} As an illustrative example, we now consider a two-level system as the target $A$, coupled to an ancillary qubit $B$. Our purpose is to construct a propagator $U$ that satisfies the condition~\eqref{mainreq} and then construct the corresponding Hamiltonian by inverse engineering~\cite{inv_Jing2013}. 

The unitary propagator of a two-level system can be generally written as
\begin{equation}\label{reciept_u}
U(t)=\cos\theta(t)+i\sin\theta(t)\vec{\sigma}\cdot\vec{n}(t),
\end{equation}
where $\vec{\sigma} = (\sigma_x,\sigma_y,\sigma_z)$ represents a vector of Pauli operators and $\vec{n}$ denotes a unit vector. The corresponding Hamiltonian can be written as \cite{Wu1980s,dq_Wu2003}
\begin{align}\label{reciept_H}
H&=i\dot{U}U^\dag \nonumber\\
&=-\vec{\sigma}\cdot[\dot{\theta}\vec{n}
+\sin\theta\cos\theta\dot{\vec{n}}
+\sin^2\theta(\dot{\vec{n}}\times\vec{n})].
\end{align}
Assume that the two-qubit unitary propagator is a direct sum of two $2\times 2$ block-diagonal matrices
\begin{equation}\label{constr_u_1}
U=\begin{pmatrix}
U_{0,g;0,g} & U_{0,g;0,e} & 0  & 0 \\
U_{0,e;0,g} & U_{0,e;0,e} & 0 & 0 \\
0 & 0 & U_{1,g;1,g} & U_{1,g;1,e} \\
0 & 0 & U_{1,e;1,g} & U_{1,e;1,e}\\
\end{pmatrix},
\end{equation}
we can directly use Eq.~\eqref{reciept_u}  to inversely engineer the blocks of $U$ and then the corresponding blocks of $H$ by Eq.~\eqref{reciept_H}. The OSMGSC conditions $U_{1,g;0,g}=U_{1,g;1,g}=0$ in~\eqref{mainreq} is accordingly satisfied. Another OSMGSC condition~$U_{1,g;1,g}=0$ in the bottom block $U_2$,
\begin{equation}\label{constr_u_2}
U_2 = \begin{pmatrix}
U_{1,g;1,g} & U_{1,g;1,e} \\
U_{1,e;1,g} & U_{1,e;1,e}
\end{pmatrix}
\end{equation}
is inversely engineered by using Eq.~\eqref{reciept_u}. Consequently the vector~$\vec{n}$ is not allowed to have the $z$ component because otherwise the component will lead to~$U_{1,g;1,g}=\cos\theta + ia\sin\theta\neq0$~($a\neq0$). For simplicity, assume that~$\vec n(t)= (- 1,0,0)$ and $\theta(t)=t$,  the $U_2$ block reads
\begin{equation}\label{constr_u_3}
U_2 = \begin{pmatrix}
\cos t & - i\sin t \\
- i\sin t & \cos t
\end{pmatrix},
\end{equation}
with $U_{1,g;1,g}=0$ at time instants $t=\pi/2 + \pi n \; ( n=0,1,2,\dots)$. The corresponding $H$ block is
\begin{equation}\label{constr_h_1}
H_2 = \begin{pmatrix}
0 & 1 \\
1 & 0 
\end{pmatrix}.
\end{equation}
Eq.~\eqref{reciept_u} does not impose constraints on $U_1$. For simplicity, we set $\theta(t)=t$, $\vec n(t)= (0,0,1)$ such that Eq.~\eqref{reciept_H} becomes
\begin{align}
U_1=\begin{pmatrix}
e^{i t} & 0 \\
0 & e^{-i t}
\end{pmatrix},  \quad
H_1=\begin{pmatrix}
-1 & 0 \\
0 & 1 
\end{pmatrix}.
\end{align}
We now combine $H_1$ and $H_2$ and write the total Hamiltonian of system $A+B$,
%\begin{equation}\label{constr_h_2}
%H = \begin{pmatrix}
%-1 & 0 & 0 & 0 \\
 %0 & 1 & 0 & 0 \\
% 0 & 0 & 0 & 1 \\
% 0 & 0 & 1 & 0 
%\end{pmatrix}
%\end{equation}
\begin{align}%\label{constr_h_3}
H&= -\ket{0,g}\bra{0,g} + \ket{0,e}\bra{0,e} \nonumber\\
&+\ket{1,e}\bra{1,g} + \ket{1,g}\bra{1,e} \nonumber\\
&=|0 \rangle\langle0|\otimes \sigma_z+|1 \rangle\langle 1|\otimes \sigma_x,\label{constr_h_3}
\end{align}
% It can also be expressed by Pauli's matrices,
% \begin{equation}\label{constr_h_3}
% H=
% \frac{1}{2}\left(I\otimes(\sigma_x - \sigma_z) - \sigma_z \otimes(\sigma_z+\sigma_x)\right)
% \end{equation}
Driven by this Hamiltonian, it is easy to verify that the system $A$ reaches its ground state by a one-shot measurement on $B$, when the measurement is made at $t=\pi/2 + \pi n$ ($n$'s are integers).

\begin{figure}
\begin{center}
\includegraphics[width=7.5cm]{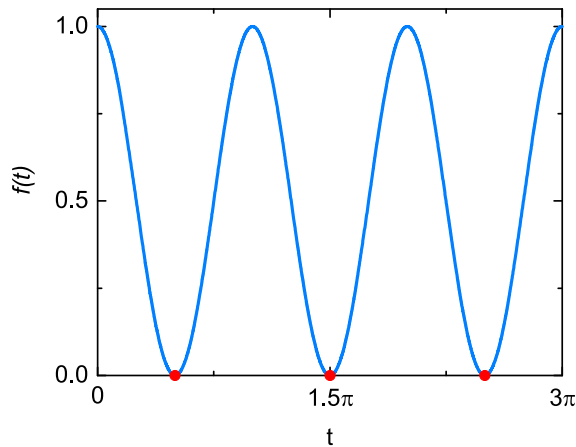}
\end{center}
\caption{(Color online) The measure $f(t)$ as a function of time. The optimal times for our one-shot projective measurements show up periodically and are highlighted by solid red circles. }%It can be observed that the one-measurement ground-state cooling condition is satisfied periodically in this system.}
\label{fig_2}
\end{figure}

To quantify the ability of achieving OSMGSC, we define a measure $f$,
\begin{equation} \label{ff}
f = \frac{1}{\sum_{m}|U_{0,g;m,g}|^2}\sum_{n\geq1, m} |U_{n,g;m,g}|^2.
\end{equation}
This measure entails that if we make measurements at times when $f=0$, the OSMGSC scheme will be fulfilled. Figure~\ref{fig_2} shows $f(t)$ for the system driven by the Hamiltonian~\eqref{constr_h_3}, where the optimal times for OSMGSC indicated by $f=0$ show up periodically over the course of time. %which shows the expected behaviours of the Hamiltonian \eqref{mainreq} . 

\begin{figure}
\begin{center}
\includegraphics[width=7.5cm]{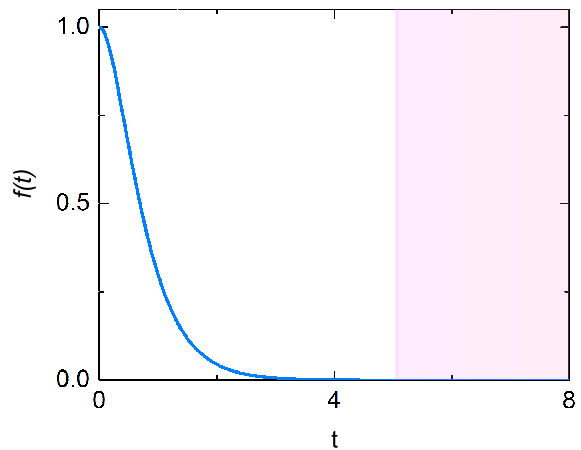}
\end{center}
\caption{(Color online) The measure $f(t)$ as a function of time. The light pink area marks the time domain where $f(t)<10^{-4}$. }
\label{fig3}
\end{figure}

We can also engineer $\theta(t)$ in $U_2$ to create a {\em steady good for cooling} state to allow our OSMGSC scheme preformed in a wide time domain. This relaxes the experimental constraints of making the measurement precisely at the optimal time instants. For example, if $\theta(t)=\pi(1-e^{-t})/2$, the corresponding Hamiltonian is
\begin{equation}
	H(t)=|0 \rangle\langle0|\otimes \sigma_z+h(t)|1 \rangle\langle 1|\otimes \sigma_x,\label{ht_dmp}
\end{equation}
where $h(t)=\dot{\theta}(t)=\pi e^{-t}/2$. Here a part of the interaction decays with time and become effectively switched off for longer time. In Fig.~\ref{fig3} we plot the measure $f(t)$ for the system under the Hamiltonian~\eqref{ht_dmp}. This Hamiltonian drives the system into a steady state $p|0,g \rangle\langle 0,g|+(1-p)|1,e \rangle\langle 1,e|$, which satisfies the OSMGSC condition~\eqref{mainreq}.

Under the constraints \eqref{mainreq}, we are free to choose $H$ and $U$ in different manners, for example by including the transition term $\ket{0}\bra{1}+\rm{h.c.}$ in the constructed Hamiltonian.  Consider the propagator $U$
\begin{equation}\label{diag_u}
U = \diag(u_1, U_2, u_3),
\end{equation}
where $u_1$, $u_3$ are complex numbers with $|u_{1}|=|u_{3}|=1$, and $U_2$ is a $2\times 2$ unitary matrix different from the previous case. By using the above method, we can obtain the corresponding $H$ with transition terms $\ket{0}\bra{1}$ and $\ket{1}\bra{0}$, 
\begin{align}\label{constr_h_5}
H =& \omega_1\ket{0,g}\bra{0,g} + \omega_2\ket{1,e}\bra{1,e}  \nonumber\\
&+\ket{1,g}\bra{0,e} + \ket{0,e}\bra{1,g},
\end{align}
where $\omega_1$ and $\omega_2$ are free parameters.

The last approach can be applied to engineering $U$ and $H$ matrices for an arbitrary $A+B$ system. Suppose that  $A$ is an $n$ level system and the ancilla $B$ is a qubit, $U$ is engineered in the same manner as~\eqref{diag_u}, {\em{i.e.}}, 
\begin{equation}
U=\diag (u_1, U_2, U_3, \dots, U_n, u_2),
\end{equation}
We then use Eqs.~\eqref{reciept_u} and ~\eqref{reciept_H} to obtain all $U_i$ ($i=2, 3, \dots, n$), where we select $\theta (t)$ and time $t$ carefully such that for each $U_i$ block, all $U_{i,g;i,g}=0$ simultaneously.

{\it Existing Hamiltonians and one-shot measurement cooling.}{\bf--} The ground-state cooling of nano mechanical resonators (NAMR) has become increasingly important in ultrahigh-precision measurements, classical to quantum transitions, preparations of nonclassical states, and quantum information processing. Efficient ground-state cooling of the NAMR with repeated measurements on an ancillary qubit has been proposed. Below we analyze the possibility of achieving OSMGSC for such a system.

When the coupling strength between the NAMR and the qubit is much smaller than the qubit frequency, the rotating wave approximation becomes valid and the total Hamiltonian is reduced to the standard Jaynes-Cummings model~\cite{Li2011,Jaynes1963},
\begin{align} \label{jc_h}
H=& \omega a^\dagger a + \frac{\Delta}{2} (\ket{e}\bra{e}-\ket{g}\bra{g})  \nonumber \\
&+g ( a\otimes\ket{e}\bra{g} + a^\dagger\otimes\ket{g}\bra{e} ),
\end{align}
where $a^\dagger$ ($a$) are creation (annihilation) operators of phonons, $\omega$ is the fundamental mode frequency of the NAMR, $\Delta$ is the tunneling amplitude between the two qubit states and $g$ is the coupling strength.  For this existing model, we study the possibility of OSMGSC by using Eq.~\eqref{mainreq}. Since the propagator $U$ has a block structure \cite{Li2011}, we can analytically give
\begin{align} \label{jc_u}
U_{0,g;0,g}(t)&=e^{i\Delta t/2}, \nonumber\\
U_{n-1,e;n-1,e}(t)&=e^{-i\varepsilon_n^+t}\cos^2\theta_n + e^{-i\varepsilon_n^-t}\sin^2\theta_n, \nonumber\\
U_{n,g;n,g}(t)&=e^{-i\varepsilon_n^+t}\sin^2\theta_n + e^{-i\varepsilon_n^-t}\cos^2\theta_n,
\end{align}
where $\varepsilon^\pm_n=\omega(n-1/2)\pm\sqrt{\left( \Delta - \omega  \right)^2 + 4g^2n}/2$ and $\tan(2\theta_n) = 2g\sqrt{n}/(\Delta-\omega)$.

\begin{figure}
\begin{center}
\includegraphics[width=7.5cm]{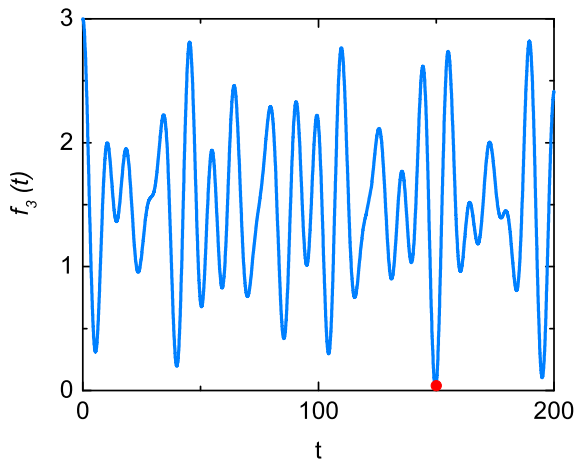}
\end{center}
\caption{(Color online) The measure  $f_3(t)$ as a function of time for the MR with $\omega=\Delta=1$ and $g=0.2$. The optimal times for our one-shot projective measurements is highlighted by a solid red circle at $t\approx 150$.}
\label{pic4}
\end{figure}

It is seen that constraints $U_{n,g;n,g}=0$ cannot be fulfilled for all $n\ge 1$ no matter how we adjust the parameters. Although the {\em{exact}} OSMGSC is impossible for this model, we may still numerically explore optimal times when the measurement should be taken. This amounts to numerical estimation of the time instants when the measure $f$ 
\begin{equation} \label{ff_jc}
f = \frac{1}{|U_{0,g;0,g}|^2}\sum_{n=1}^\infty |U_{n,g;n,g}|^2,
\end{equation}
approximates to zero, though not completely vanishes.

If we consider a thermal input state as in the conventional cooling process, 
\begin{equation} \label{term_state}
\rho_A^i = \frac{1}{Z}\sum_{n=0}e^{-n\omega/T}\ket{n}\bra{n}, \; Z=\sum_{n=0}e^{-n\omega/T},
\end{equation}
our numerical analysis shows that only the first few blocks of $U$ determine the possibility of OSMGSC when $T$ is not too large. Therefore, we can truncate the sum to the $k$-th level,
\begin{equation} \label{fk_jc}
f_k = \frac{1}{|U_{0,g;0,g}|^2}\sum_{n=1}^k |U_{n,g;n,g}|^2.
\end{equation}
Using Eqs.~\eqref{p_n_g}, ~\eqref{term_state} and~\eqref{fk_jc} we can derive an inequality for the cooling success probability $P_{c}$
\begin{equation} \label{ineq}
P_{c} = \frac{P_{0,g}}{P_g} > 1 - e^{-\omega/T}\left( f_k + Z^2e^{-k\omega/T} \right),
\end{equation}
where $P_g=\sum_{n=0}^{\infty}P_{n,g}$ is the probability of obtaining the qubit ground state $\ket{g}$.
We illustrate $f_k$ in a resonate case with $\omega=\Delta=1$ and $g=0.2$. Figure~\ref{pic4} shows $f_3(t)$ for the parameters given in the figure caption. Interestingly, we indeed find $f_3(150)=0.04\approx 0$, which is the optimal time to implement a projective measurement to achieve the ground-state cooling of the NAMR. The inequality~\eqref{ineq} also concludes that probability of achieving the ground-state cooling is greater than 93\% at $T=1$. The example suggests the possibility that the propagator $U$ generated by certain existing Hamiltonians may directly result in OSMGSC at specific time instants or domains. 

As another example, Ref.~\cite{Xu2014} uses a specific joint unitary matrix $U$,  an ordered product of a Hadamar gate, a phase $\gamma$ shifter on ancillary qubit $B$, a controlled--\,$\mathcal{U}(s)$ and another Hadamar gate on $B$. Here $\mathcal{U}(s)=\exp(-i\mathcal{H}s)$, where $\mathcal{H}$ is the Hamiltonian of the target system $A$. A direct calculation shows that $U_{1,g;0,g}=0$ and $U_{1,g;1,g}(s)=1-i\exp(i\gamma)[\cos s - i\sin s]$. Interestingly, our requirement~\eqref{mainreq} can be met when $s=\pi/2$ and  $\gamma=0$. Our OSMGSC scheme helps to find the optimal measurement instant of the experiment.

{\it Conclusion.}{\bf--} In conclusion, we proved the existence of a family of systems that can be efficiently cooled to their ground states by making one-shot projective measurement on a coupled ancilla. The explicit condition for achieving this one-measurement cooling is given, and we also show a general procedure for finding the corresponding Hamiltonian to realize this technique by means of inverse engineering. For existing Hamiltonians, our method can be used to find an optimal time when the projective measurement is taken. Our approach provides clear guidance for experimental endeavours in rapid cooling techniques.% and it is found that we can still have a high rate of successof cooling even if the system does not strictly meet the one-measurement cooling condition.

%\bibliographystyle{prs}
%\bibliography{gsc1}

%%%%%%%%%%%%%%%%%%%%%%%%%%%%%%%%%%%%
%\begin{thebibliography}{99}

%\end{thebibliography}

\end{document}